\documentclass{aa}

\usepackage{natbib}
\bibpunct{(}{)}{;}{a}{}{,} 
\usepackage{amssymb}
\usepackage{url}
\usepackage{hyperref}
\usepackage{graphicx}
\usepackage{longtable}
\usepackage[usenames,dvipsnames]{color}
\usepackage{booktabs}
\usepackage{lscape}

\newcommand{\submm}{submillimetre}

\newcommand{\micron}{\mbox{$\mu$m}}
\newcommand{\msun}{\mbox{$M_\odot$}}
\newcommand{\mstar}{\mbox{$M_*$}}
\newcommand{\mdust}{\mbox{$M_{\rm dust}$}}

\newcommand{\kms}{\mbox{km\,s$^{-1}$}}

\urldef{\hyperleda}\url{http://leda.univ-lyon1.fr/ledacat.cgi?NGC%203278}

\newcommand{\magphys}{\sc Magphys}

\begin{document}

 \title{The fate of the interstellar medium in early-type galaxies. I. First direct measurement of the timescale of dust removal
 }
 
\titlerunning{The ISM in ETGs I: measurement of the timescale of dust removal}
\authorrunning{Micha{\l}owski et al.}

\author{Micha{\l}~J.~Micha{\l}owski\inst{\ref{inst:poz},\ref{inst:roe}}
\and
J.~Hjorth\inst{\ref{inst:dark}}
\and
C.~Gall\inst{\ref{inst:dark}}
\and
D.~T.~Frayer\inst{\ref{inst:nrao}}
\and
A.-L.~Tsai\inst{\ref{inst:poz}}
\and
H.~Hirashita\inst{\ref{inst:asiaa}}
\and
K.~Rowlands\inst{\ref{inst:jhu},\ref{inst:stand}}
\and
T.~T.~Takeuchi\inst{\ref{inst:nagoya}}
\and
A.~Le\'{s}niewska\inst{\ref{inst:poz}}
\and
D.~Behrendt\inst{\ref{inst:roe}}
\and
N.~Bourne\inst{\ref{inst:roe}}
\and
D.~H.~Hughes\inst{\ref{inst:inaoe}}
\and
E.~Spring\inst{\ref{inst:roe}, \ref{inst:ams}}
\and
J.~Zavala\inst{\ref{inst:inaoe},\ref{inst:ut}}
\and
P.~Bartczak\inst{\ref{inst:poz}}
        }

\institute{
Astronomical Observatory Institute, Faculty of Physics, Adam Mickiewicz University, ul.~S{\l}oneczna 36, 60-286 Pozna{\'n}, Poland, {\tt mj.michalowski@gmail.com}\label{inst:poz}
\and
SUPA\thanks{Scottish Universities Physics Alliance}, Institute for Astronomy, University of Edinburgh, Royal Observatory, Blackford Hill, Edinburgh, EH9 3HJ, UK  
\label{inst:roe}
\and
DARK, Niels Bohr Institute, University of Copenhagen, Lyngbyvej 2, DK-2100 Copenhagen \O, Denmark  
\label{inst:dark}
\and
National Radio Astronomy Observatory, P.O. Box 2, Green Bank, WV 24944, USA \label{inst:nrao}
\and
Institute of Astronomy and Astrophysics, Academia Sinica, Astronomy-Mathematics Building, AS/NTU, No. 1, Sec. 4, Roosevelt Road, Taipei 10617, Taiwan \label{inst:asiaa}
\and
Department of Physics \& Astronomy, Johns Hopkins University, Bloomberg Centre, 3400 N. Charles Str, Baltimore, MD 21218, USA \label{inst:jhu}
\and
SUPA$^\star$, School of Physics \& Astronomy, University of St Andrews, North Haugh, St Andrews, Fife KY16 9SS, UK \label{inst:stand}
\and
Division of Particle and Astrophysical Science, Nagoya University, Furo-Cho, Chikusa-ku, Nagoya 464-8602, Japan \label{inst:nagoya}
\and
Instituto Nacional de Astrof\'{\i}sica, \'Optica y Electr\'onica (INAOE), Aptdo. Postal 51 y 216, 72000 Puebla, Pue., Mexico\label{inst:inaoe}
\and
Anton Pannekoek Institute, University of Amsterdam, Science Park 904, NL-1098 XH Amsterdam, the Netherlands\label{inst:ams}
\and
The University of Texas at Austin, 2515 Speedway Blvd Stop C1400, Austin, TX 78712, USA \label{inst:ut}
}

\abstract{
An important aspect of quenching star formation is the removal of the cold interstellar medium (ISM; non-ionised gas and dust) from a galaxy.
In addition, dust grains can be destroyed in a hot or turbulent medium.
The adopted timescale of dust removal usually relies on uncertain theoretical estimates.
 It is tricky to track dust removal because the  dust is constantly being replenished by consecutive generations of stars. 
}
{Our objective is to carry out an observational measurement of the timescale of dust removal.
}
{We explored an approach 
to select galaxies that demonstrate detectable amounts of dust and cold ISM coupled with a low current dust production rate. 
Any decrease of the dust and gas content as a function of the age of such galaxies  must, therefore, be attributed to processes governing ISM removal.
We used a sample of the galaxies detected by {\it Herschel} in the far-infrared with visually assigned early-type morphology or spirals with red colours. We also obtained JCMT/SCUBA-2 observations for five of these galaxies.
}
{We discovered an exponential decline of the dust-to-stellar mass ratio with age, which we interpret as an evolutionary trend for the dust removal of these galaxies.
For the first time, we have directly measured the dust removal timescale for such galaxies, with a result of $\tau=(2.5\pm0.4)\,$Gyr (the corresponding half-life time is $(1.75\pm0.25)$ Gyr).
This quantity may be applied to models in which it must be assumed {\em a priori} and cannot be derived.
}
{Any process which removes dust in these galaxies, such as dust grain destruction, cannot happen on shorter timescales.
The timescale is comparable to the quenching timescales found in simulations for galaxies with similar stellar masses.  
The dust is likely of internal, not external origin. It was either formed in the past directly by supernovae (SNe) or from seeds produced by SNe, and with grain growth in the ISM contributing substantially to the dust mass accumulation.
}


\maketitle

\section{Introduction}
\label{sec:intro}

An important aspect of quenching star formation is the removal of the cold interstellar medium (ISM; non-ionised gas and dust) from a galaxy.
This can happen by outflows, astration, or  ionisation of gas by stars or active galactic nuclei. In addition, dust grains can be destroyed in a hot or turbulent medium by sputtering and sublimation \citep{barlow78,dwek96,silvia10,slavin15}.
This process must be taken into account when modelling dust evolution in nearby and distant 
galaxies \citep[e.g.][]{gall11b,lagos14b,lagos15,zhukovska16,krumholz17}. 
Dust is an important constituent of the cold ISM and can be used as its tracer  \citep[e.g.][]{scoville14,scoville16}. 

The adopted timescale of dust removal usually relies on uncertain theoretical estimates.
An efficient process involves supernova (SN) forward shocks sweeping up dusty ISM material and, thereby, 
completely destroying existing dust grains \citep[e.g.][]{jones11}. 
Depending on the environment in the ISM, as well as on the assumed dust grain properties and  
destruction mechanisms, the dust destruction timescale may be as short as a few 
thousand years 
up to $\sim1\,$Gyr 
\citep{draine79,jones94,jones04,zhukovska16}. 

Observationally, estimating dust 
destruction rates in galaxies is impacted by substantial uncertainties. Recent 
investigations of SN remnants in the Milky Way and the Magellanic Clouds suggest a short destruction timescale of a few tens of Myr \citep{temim15,lakicevic15}.
However, for such short timescales it would be very difficult to build up and 
maintain significant amounts of dust in any galaxy in the absence of a very 
efficient dust production mechanism which would constantly replenish the dust to 
counter its rapid destruction \citep[e.g.][]{gall11, gall18}.
On the other hand, \citet{barlow78} and \citet{slavin15} show that dust 
destruction due to SN forward shocks, mainly by sputtering, should occur on a timescale of $1$--$3$\, Gyr 
(see \citealt{gall18} for a recent discussion). However, the complete destruction of 
dust grains through SN shocks may not be the only mechanism at play in the removal of 
dust from a galaxy. Heating by planetary nebulae, cosmic rays, X-rays, or other high energy radiation can also alter dust grains. In addition, dust may be blown out of galaxies due to the feedback of stars and SNe 
\citep[e.g.][]{finkelstein12,conroy15,spilker18,jones19,li19b}

The vast amounts of molecular gas found in post-starburst galaxies indicate that the process of removing cold ISM after the starburst (e.g. through outflows, AGN or stellar feedback) is not rapid \citep{rowlands15,french15,alatalo16,smercina18}.
\citet{rowlands15} analyse a sample of starburst and post-starburst galaxies at ages up to $\sim1$\,Gyr after the starburst and do not find any clear decrease of the molecular gas content with age, indicating that ISM removal must operate on longer timescales. They did find a weak decreasing trend of dust mass with age, consistent with dust removal during the first $\sim200$\,Myr after the starburst. The dust temperature exhibits a similar decrease. 
On the other hand, combining all post-starburst galaxies with molecular gas measurements, \citet{french18} and \citet{li19} detect a weak anti-correlation of the molecular gas fraction with age, deriving a gas removal timescale of $\sim100$--$200$\,Myr.

In contrast, the only observational estimate of the lifetimes of dust grains in the hot ISM of early-type galaxies (ETGs) 
turned out to be very short, $<(45\pm25)$\,Myr, based on dust production rates derived for asymptotic giant branch (AGB) stars from mid-infrared observations and non-detections by {\it Herschel} \citep{clemens10}. 
The galaxies were selected based on their lack of any sign of recent star-formation in the mid-infrared spectra, so they most likely represent the oldest early-type galaxies. 
Moreover, these galaxies are located in the Virgo cluster, so this timescale may apply to cluster-specific processes.

The analysis of dust removal in galaxies is complicated by the fact that most of the studies on this subject have concentrated on star-forming spiral/irregular galaxies \citep{draine07,draine09,temim15,lakicevic15} in which both the dust destruction rate is and the dust production rate are high, due to on-going star formation.
Hence, it is tricky to track the dust removal, because usually the dust is constantly being replenished by consecutive generations of stars. 
Here we explore an alternative approach, that is,
through the selection of those   galaxies which demonstrate detectable amounts of dust and cold ISM while exhibiting
a low current SN dust production rate because of the extensive amount of time elapsed since the last major star-formation event.
This can be achieved by selecting dusty early-type galaxies (with little star formation).
Therefore, any decrease of dust and gas content as a function of the age of such galaxies must be attributed to processes governing the ISM removal.
This selection obviously limits our ability to study dust destruction processes connected with SNe (as their rate is also low), but allows us to explore dust removal in galaxies on the way to becoming passive.

With large-area {\it Herschel} surveys, the dust emission of significant numbers of such (rare) dusty early-type galaxies have been detected \citep{bourne12,rowlands12,smith12,agius13,agius15,dariush16}.
Here we use the term ETG for galaxies morphologically classified as ellipticals, lenticulars, or early-type spirals (Sa and SBa).
In this series of papers, we embark on a study of the ISM removal in such early-type galaxies. 

The objective for this paper is 
to carry out an observational measure of  the timescale of dust removal.
In future papers of this series, we intend to constrain the physical mechanism of this process based on the gas properties of these galaxies and theoretical modelling.

We use a cosmological model with $H_0=70$ km s$^{-1}$ Mpc$^{-1}$,  $\Omega_\Lambda=0.7$, and $\Omega_m=0.3$. 
All calculations are carried out assuming the \citet{chabrier03} initial mass function (IMF).

\section{Sample of dusty early-type galaxies}
\label{sec:sample}

We use the sample of dusty early-type galaxies of \citet{rowlands12}. This sample includes all galaxies with elliptical morphology and red spirals in the {\it Herschel}  Astrophysical Terahertz Large Area Survey  \citep[H-ATLAS;][]{hatlas} $\sim14\,\mbox{deg}^2$ Science Demonstration Field \citep{ibar10,pascale11,rigby11,smith11} that are detected at $250\,\mu$m (all of them are also detected at $350\,\mu$m and most of them at $500\,\mu$m). 
The specific selection criteria used by \citet{rowlands12} are:
\begin{enumerate}
\item Set within the H-ATLAS Science Demonstration Field (solely positional).
\item Matched to an optical Sloan Digital Sky Survey (SDSS) source with a spectroscopic redshift 
(of a redshift range $0.01<z_{\rm spec}<0.32$)  
within a $10''$ radius and with $>0.8$ reliability (removes high-$z$ galaxies).
\item {\it Herschel} $250\,\mu$m $>5\sigma$ detection (selects dusty galaxies).
\item Visually classified as early-type (elliptical or S0), or red spiral with near-ultraviolet (NUV) to $r$-band colour of $\mbox{NUV} -r > 4.5$ (morphology and colour).
\end{enumerate}

We note that, in principle, criterion 2 does not remove high-$z$ dusty galaxies lensed by nearby ones \citep{negrello10,negrello17,bussmann13,wardlow13}. However, the maximum $500\,\micron$ flux of the sources in the sample of \citet{rowlands12} is 62\,mJy, and 90\% of them are below 30
\,mJy. At such fluxes the contribution of lensed sources is expected to be, at most, only a few percent \citep{negrello10,negrello17,wardlow13}.

The sample consists of 61 galaxies, including 42 ellipticals or lenticulars, and 19 red spirals, most of which are classified as Sa or SBa galaxies. These galaxies will be referred to here as ETGs. 
\citet{rowlands12} fitted the ultraviolet-to-submm spectral energy distribution [from the Galaxy And Mass Assembly (GAMA) survey; \citealt{driver11,driver16,hill11,robotham10,baldry10}] using Multi-wavelength Analysis of Galaxy Physical Properties  \citep[{\magphys};][]{dacunha08}. {\magphys} fits the photometry to account for the balance between energy absorbed by dust and re-emitted in the infrared with the dust that is distributed through two components: diffuse and cloud-like.  
In particular, \citet{rowlands12} present {\magphys}-based estimates of stellar and dust (with an opacity coefficient of $\kappa_{250\,\micron}=0.89\,\mbox{m}^2\,\mbox{kg}^{-1}$) masses, star formation rates (SFRs), and  luminosity-weighted ages of stellar populations, which are  used here. \citet{rowlands12} confirm the passive nature of most of these galaxies by measuring low SFRs and specific SFRs (normalised by stellar mass), which are well below those of galaxies on the star-formation main sequence \citep[also see: ][]{hunt14,hjorth14}, and a  long time having passed ($>1\,$Gyr) since the latest star-formation episode.

\begin{figure*}
\includegraphics[width=0.9\textwidth]{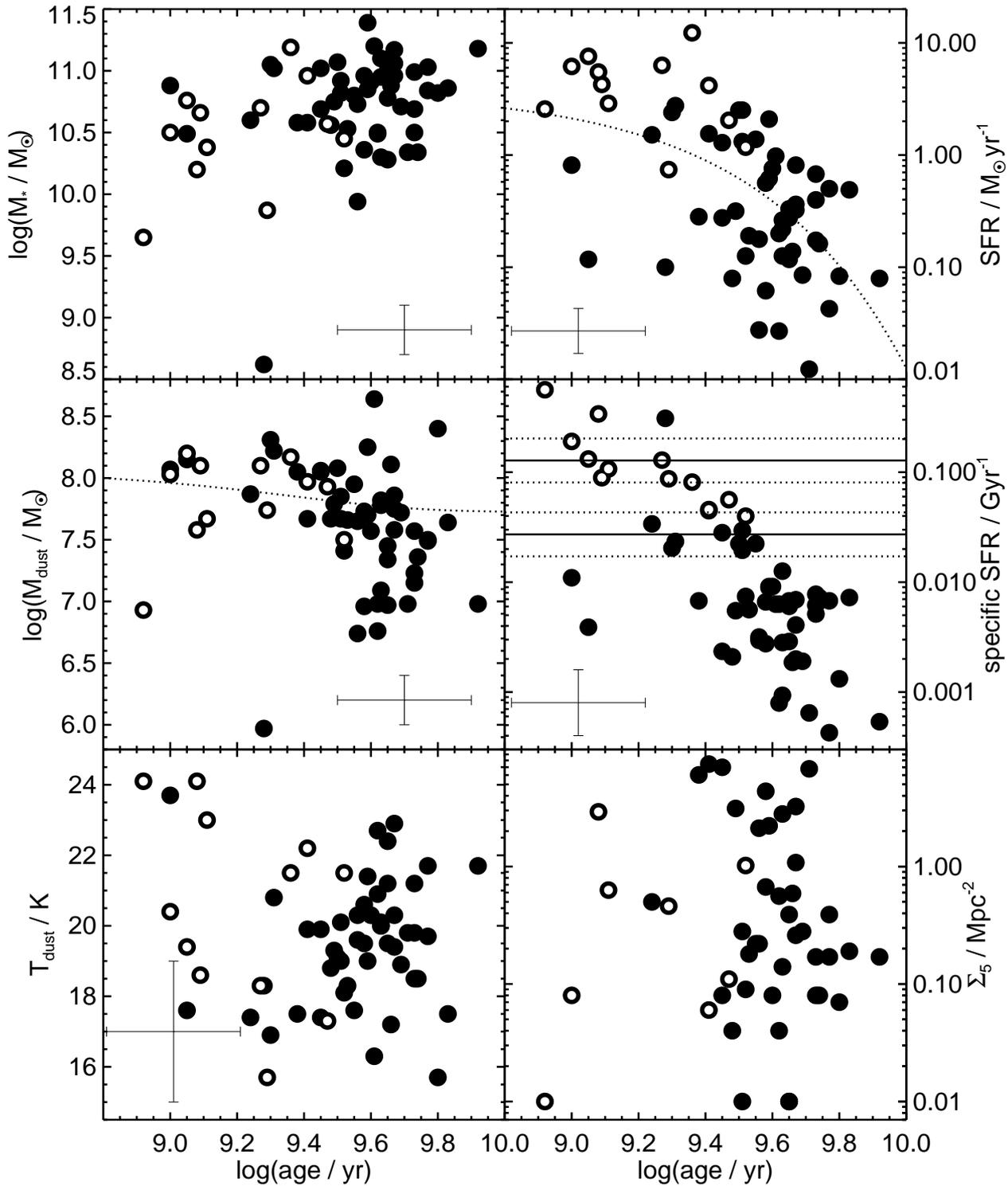}
 \caption{Stellar mass ({\it top left}), SFR ({\it top right}), dust mass ({\it middle left}), specific SFR ({\it middle right}), temperature of  cold diffuse ISM dust component ({\it bottom left}) and  galaxy number surface density ({\it bottom right}) derived by \citet{rowlands12} as function of luminosity-weighted stellar age. The typical errors are shown as large crosses.   On the SFR panel, the dotted line{\it } shows the exponential fit to the data, whereas on the {\mdust}  panel, the dotted line denotes the evolution of dust mass affected only by astration resulting from this fit and the assumption of the gas-to-dust mass ratio of 100. A higher ratio would result in an even flatter evolution.
 The specific SFR panel shows the star formation main sequence at $z=0.13$ for stellar masses of $\log(M_*/\msun)=10$ (higher line) and $11.5$ (lower line)  as  solid lines, with the scatter shown as  dotted lines \citep{speagle14}.
 The galaxy number surface density is defined as $\Sigma_5=5/\pi d_5^2$, where $d_5$ is  the projected comoving distance to the 5th nearest neighbour within $\pm1000\,\kms$. 
 Open circles denote galaxies which are within the main sequence (see Fig.~\ref{fig:sfr_ms}). Filled symbols which appear to be within the main sequence are galaxies at higher redshifts, for which the main-sequence is higher.
}
 \label{fig:all_age}
 \label{fig:tdust}
 \label{fig:sfr_age}
 \label{fig:ssfr_age}
 \label{fig:den}
 \label{fig:ms_age}
 \label{fig:md_age}
\end{figure*}

\begin{figure*}
\includegraphics[width=\textwidth]{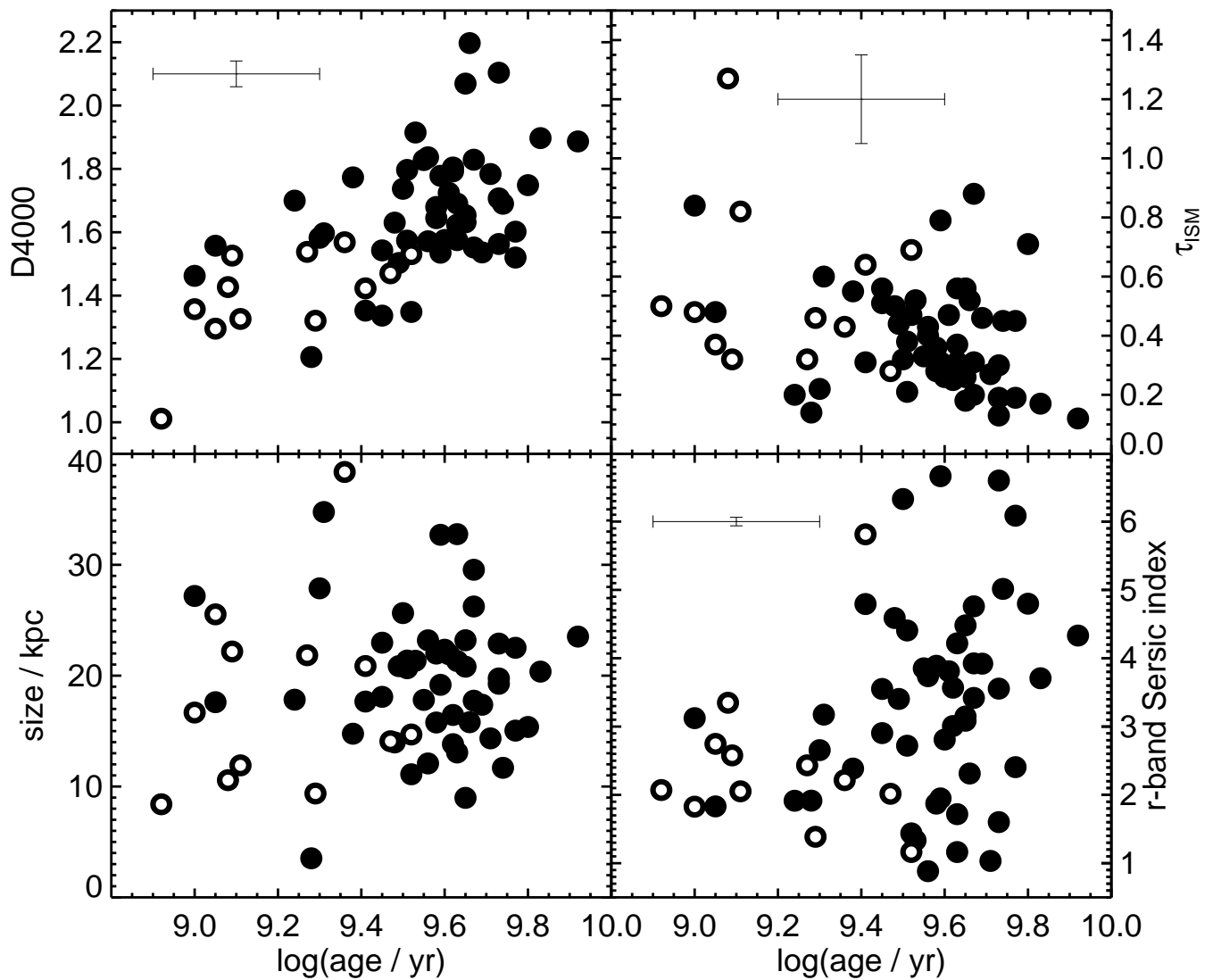}
 \caption{
 D4000 spectral index measured from SDSS fiber spectroscopy. ({\it top left}),
 optical depth 
 in the diffuse ISM ({\it top right}) in the {\magphys} model derived by \citet{rowlands12}, isophotal $r$-band major axis ({\it bottom left}), and $r$-band Sersic index  ({\it bottom right}; \citealt{kelvin12}) as a function of luminosity-weighted stellar age. The typical errors are shown as large crosses.}
 \label{fig:all_age2}
 \label{fig:taumc}
 \label{fig:d4000}
 \label{fig:tauism}
 \label{fig:size}
 \label{fig:sersic}
\end{figure*}

\begin{figure}
\includegraphics[width=0.48\textwidth]{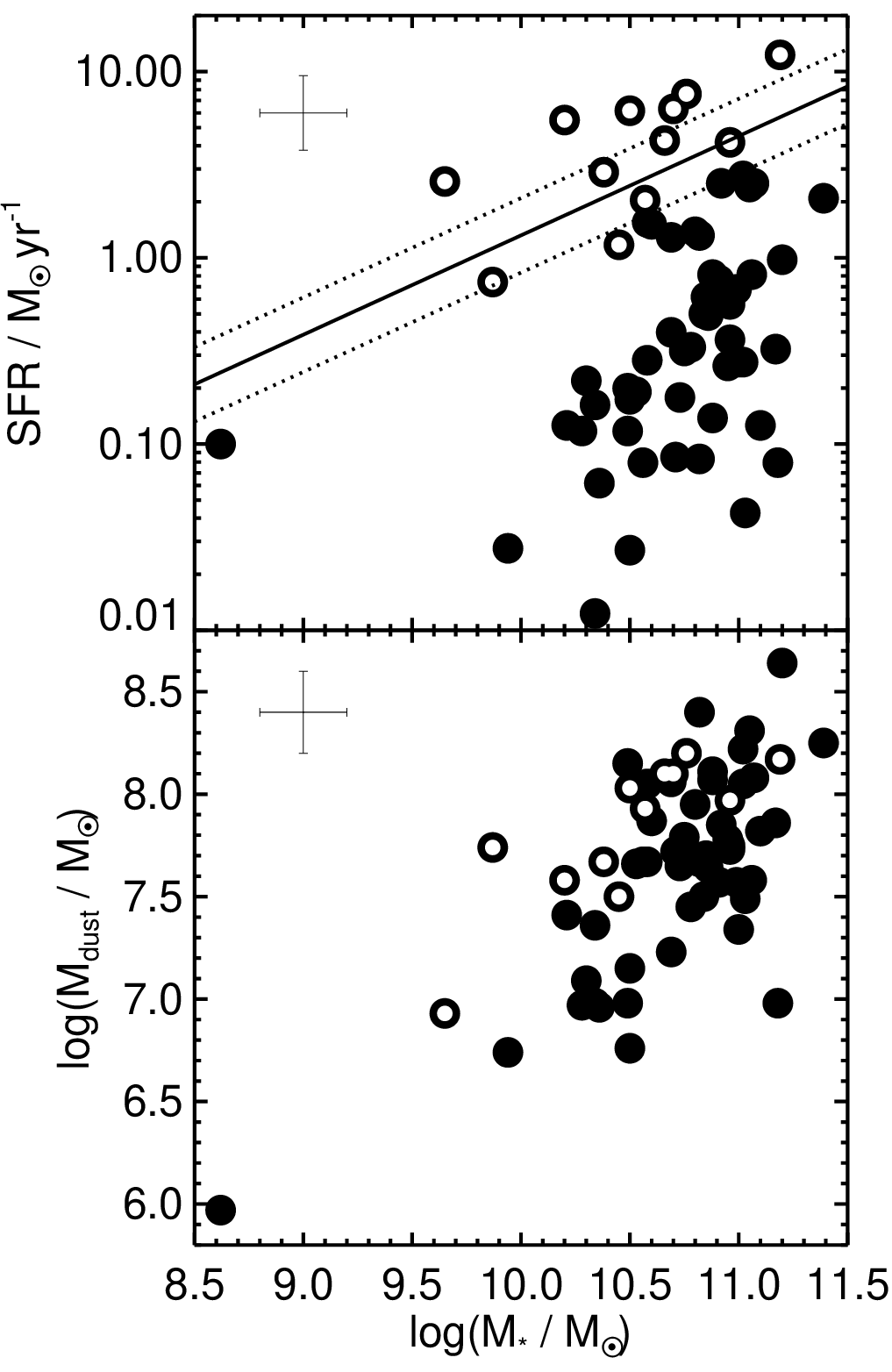}
 \caption{SFR ({\it top}) and dust mass ({\it bottom}) as function of stellar mass derived by \citet{rowlands12}. 
 The star formation main sequence at $z=0.13$ is plotted as a solid line, with the scatter shown as  dotted lines \citep{speagle14}. The early-type galaxies within the main sequence are plotted as open symbols. Filled symbols which appear to be within the main sequence are galaxies at higher redshifts, for which the main-sequence is higher.
 The typical errors are shown as large crosses. The Spearman rank correlation coefficient for the $M_d$-$M_*$ diagram is 0.5 with a very small probability ($\sim3\times10^{-5}$, $\sim4\sigma$) of the null hypothesis (no correlation) being acceptable.
}
\label{fig:all_ms}
\label{fig:sfr_ms}
 \label{fig:md_ms}
\end{figure}

The basic properties of the galaxies in our sample are shown in Figs.~\ref{fig:all_age}, \ref{fig:all_age2}, and \ref{fig:all_ms}.
These figures show which properties evolve with age and which are relatively constant.
The Sersic indices in the $r$-band of our galaxies are measured by \citet{kelvin12} within the GAMA Data Release 3 \citep{baldry18}.
We also show the star-formation main sequence  on the relevant panels \citep{speagle14} and mark galaxies that are located within the main sequence with empty symbols.

\section{JCMT/SCUBA-2 observations}
\label{sec:data}

In order to assess the robustness of the {\it Herschel}-only dust mass estimates,
we obtained longer wavelength data for some of these galaxies. These data would allow us to detect very cold dust, which, in
principle, could dominate the total dust mass budget. 

We obtained $850\,\micron$ and $450\,\micron$ data using the James Clerk Maxwell Telescope (JCMT) equipped with the Submillimetre Common-User Bolometer Array 2 \citep[SCUBA-2;][]{scuba2} for a subset of the galaxies with higher ages  ($>2$\,Gyr; proposals M15AI116 and M15BI010, PI: Micha{\l}owski).
We utilised the standard `daisy' scanning pattern, suitable for compact sources.
The data were retrieved from the Canadian Astronomy Data Centre (CADC) and were  reduced with the standard pipeline in {\sc Smurf}\footnote{\url{www.starlink.ac.uk/docs/sun258.htx/sun258.html}} package 
\citep{chapin13} with the flux calibration factor (FCF) of 
537 Jy pW$^{-1}$ beam$^{-1}$ for 850\,{\micron} and 491 Jy pW$^{-1}$ beam$^{-1}$ for 450\,{\micron}
\citep{dempsey13}. The full width at half maximum (FWHM) of the resulting $850$ and $450\,\micron$ maps is 14.6 and 7.9\,arcsec, respectively.
 The observing log 
 is presented in Table~\ref{tab:scuba2}.

\begin{figure*}
\includegraphics[width=\textwidth]{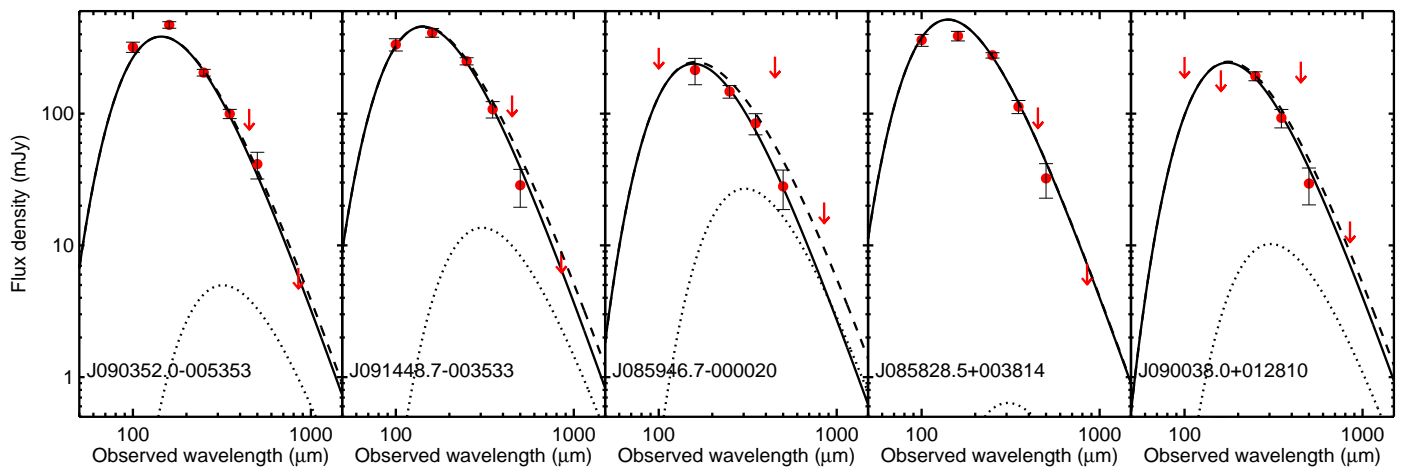}
 \caption{Spectral energy distribution of  sub-sample observed by JCMT/SCUBA-2. Circles and arrows denote detections and $3\sigma$ upper limits, respectively. The solid lines are the grey-body models using the diffuse ISM temperature derived by \citet{rowlands12} and the emissivity index $\beta=2$. In three cases, the $850\,\mu$m data are deep enough to rule out contribution of very cold dust. Dotted lines denote the models of very cold dust with the temperature of 10\,K, whereas the dashed lines denote the sum of both solid and dotted curves. The very cold dust model was scaled so that the total emission matches the SCUBA-2 upper limits, or {\it Herschel} data, if they are more constraining.
 The panels are ordered from left to right according to their increasing ages and these galaxies are marked in Fig.~\ref{fig:mdms_age}.
}
 \label{fig:sed}
\end{figure*}

\begin{table*}
\caption{Observation log and results for JCMT/SCUBA-2 observations.}
\label{tab:scuba2}
\begin{center}
\small
\begin{tabular}{llcccccc}
\hline\hline
Galaxy          & Obs. date     & $t_{\rm int}$ & $\tau_{\rm CSO,225\,GHz}$     & $F_{850}$ & $F_{450}$ & T$_{\rm dust}$ & $M_{\rm cold}/$                \\
                        &                       &  (hr)         &               & (mJy)           & (mJy) & (K) & $M_{\rm Herschel}$  \\
\hline
J090352.0$-$005353 & 2015 Apr 27, May 19, 2016 Jan 07 & 1.5             & 0.07--0.075, 0.05, 0.03 &       $0.7\pm 2.0$  & $-11\pm 40$ & 22.2 & 0.4 \\
J091448.7$-$003533  & 2015 Apr 21 & 1.0                 & 0.023--0.051 &      $-3.1\pm4$ & $47\pm       30$ & 21.5 & 0.8 \\
J085946.7$-$000020 & 2015 Dec 17, 2016 Jan 05 & 1.0          & 0.04, 0.03 &  $0.2\pm7$ & $-29\pm100$ & 19.5 & 1.8\\
J085828.5+003814 & 2015 Apr 21 & 1.0            & 0.044--0.047 &  $-1.7\pm 3.0$ & $21\pm30$ & 21.7 &  0.04\\
J090038.0+012810 & 2016 Jan 05, 07 & 1.0          & 0.03 &  $0.2\pm5$ & $68\pm60$ & 17.5 & 0.4\\

\hline 
\end{tabular}
\tablefoot{The columns are: galaxy name, observation date, on-source integration time, 225\,GHz opacity measured by Caltech Submillimeter Observatory (CSO), measured flux at 850 and 450\,{\micron}, ISM dust temperature derived from the {\it Herschel} data \citep{rowlands12}, and the ratio of the maximal very cold dust with the temperature of 10\,K allowed by the data to the dust mass measured with {\magphys}.
}
\end{center}
\end{table*}

For each galaxy observed by JCMT/SCUBA-2, 
we performed aperture photometry on the $850\,\micron$ and $450\,\micron$ maps using  aperture sizes equal to the SDSS DR7 isophotal major axis, which was also used for {\it Herschel} aperture fluxes \citep{pascale11,rigby11}. The noise was assessed by placing apertures of the same size at random around the target position. We did not detect any of the targets.
The measured fluxes are presented in Table~\ref{tab:scuba2}.

The spectral energy distributions including {\it Herschel} data are shown in Fig.~\ref{fig:sed}.
We overplot the grey-body models using the cold-dust temperatures derived by \citet{rowlands12} and the emissivity index $\beta=2$.
In three out of five cases, our upper limits are very close to the extrapolations from shorter wavelengths, ruling out the existence of a {\submm} excess, which could be the manifestation of very cold dust.

In order to assess the amount of very cold dust allowed by the data, we add a grey-body model with a temperature of 10\,K (below the coldest average dust temperature ever reported for a galaxy; e.g.~\citealt{chapman05,michalowski08,symeonidis13}). If we assume a higher temperature, we would infer an even lower cold dust mass. We scale this model so that its sum with the model derived using {\it Herschel} data matches the SCUBA-2 upper limits. For two galaxies with shallower SCUBA-2 limits (J085946.7-000020 and J090038.0+012810) this causes the total model to overpredict the {\it Herschel} fluxes, so we scale the very cold model further down so that the total model matches the {\it Herschel} fluxes. This allowed us to measure the limits on the dust masses of this component. Their ratios to the {\it Herschel}-derived masses are shown in the last column of Table~\ref{tab:scuba2}. In 
all
cases, the very cold dust component has, at most, twice the mass of the warmer dust. Even if such amounts of very cold dust are indeed present in galaxies with high ages, this effect is not sufficiently strong to explain the  {\mdust} drop of 1.5\,dex shown of Fig.~\ref{fig:md_age}.

\section{A declining dust mass with age}
\label{sec:res}

\begin{figure*}
\includegraphics[width=0.8\textwidth]{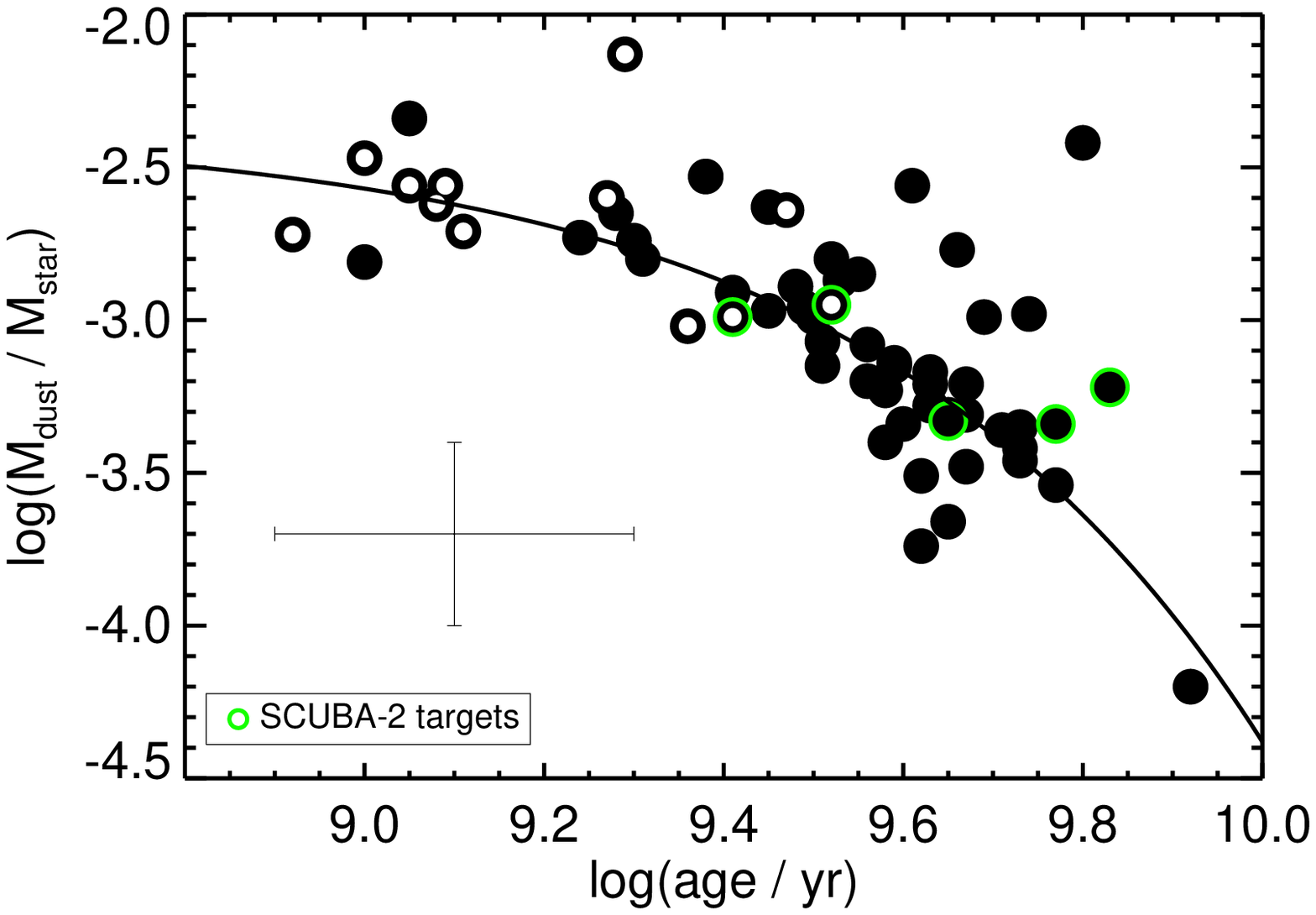}
 \caption{Dust-to-stellar mass ratio as function of luminosity-weighted stellar age of early-type galaxies detected by {\it Herschel}  \citep[{\it }black circles;][]{rowlands12}.  The exponential fit (solid line) has the  lifetime $\tau=(2.5\pm0.4)\,$Gyr and the corresponding half-life time of $\sim(1.75\pm0.25)$ Gyr. The typical errors  (large cross,  $\sim0.3$ dex for $\mdust/\mstar$ and $\sim0.2$ dex for the age) are  much smaller than the range of the values.  
The targets of JCMT/SCUBA-2 observations are marked as green circles. Open circles denote galaxies which are within the main sequence (see Fig.~\ref{fig:sfr_ms}).
}
 \label{fig:mdms_age}
\end{figure*}

\begin{figure*}
\includegraphics[width=0.8\textwidth]{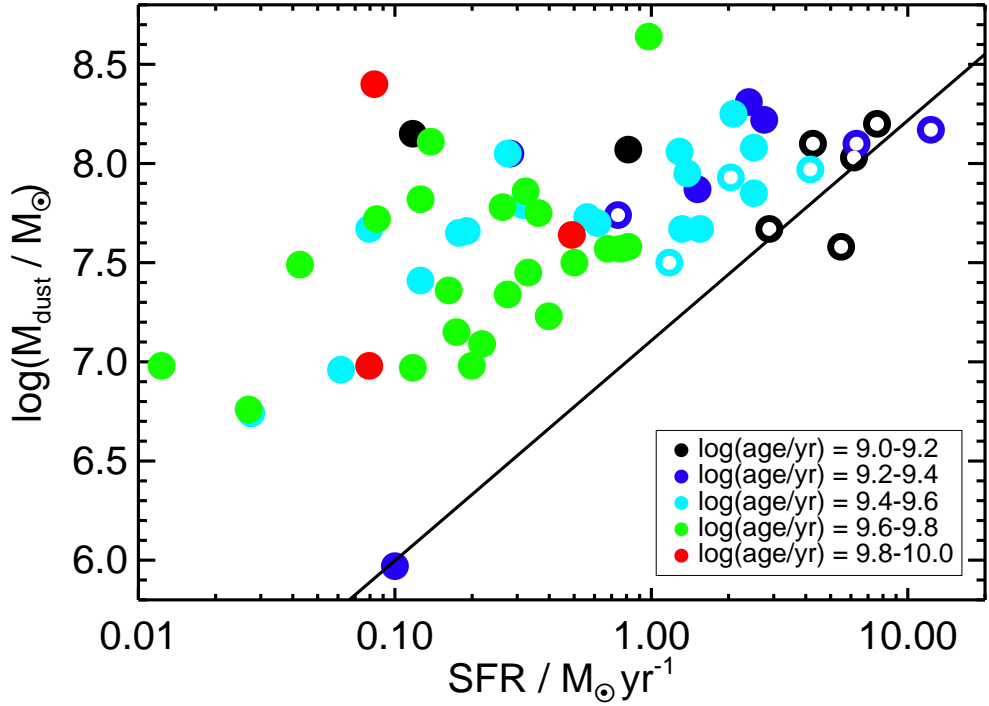}
 \caption{Dust mass as function of the star formation rate of galaxies in our sample (circles) and colour-coded by age. The solid line denotes the relation for star-forming galaxies derived by \citet{dacunha10}. Dusty early-type galaxies are moving away from this relation diagonally towards bottom-left, as predicted by \citet{hjorth14}. Open circles denote galaxies which are within the main sequence (see Fig.~\ref{fig:sfr_ms}).
}
 \label{fig:md_sfr}
\end{figure*}

\begin{figure*}
\includegraphics[width=0.8\textwidth]{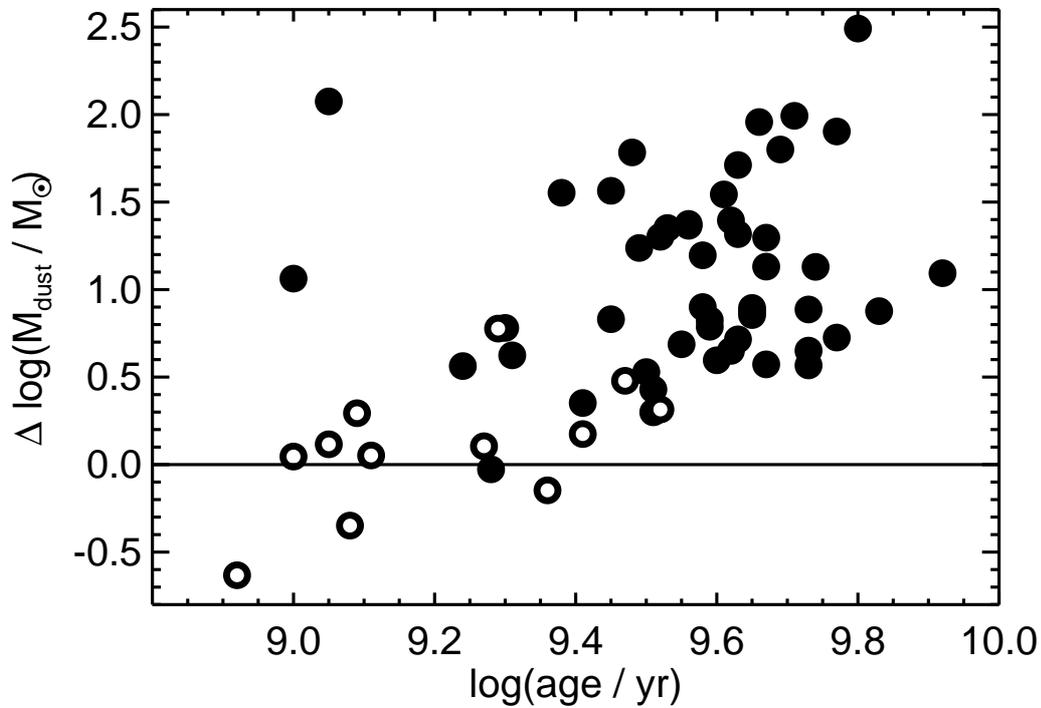}
 \caption{Distance from $\mdust$-SFR relation of \citet[][Fig.~\ref{fig:md_sfr}]{dacunha10} as function of luminosity-weighted stellar age for galaxies in our sample. Open circles denote galaxies which are within the main sequence (see Fig.~\ref{fig:sfr_ms}).
}
 \label{fig:mdsfr_age}
\end{figure*}

\begin{figure}
\includegraphics[width=0.48\textwidth]{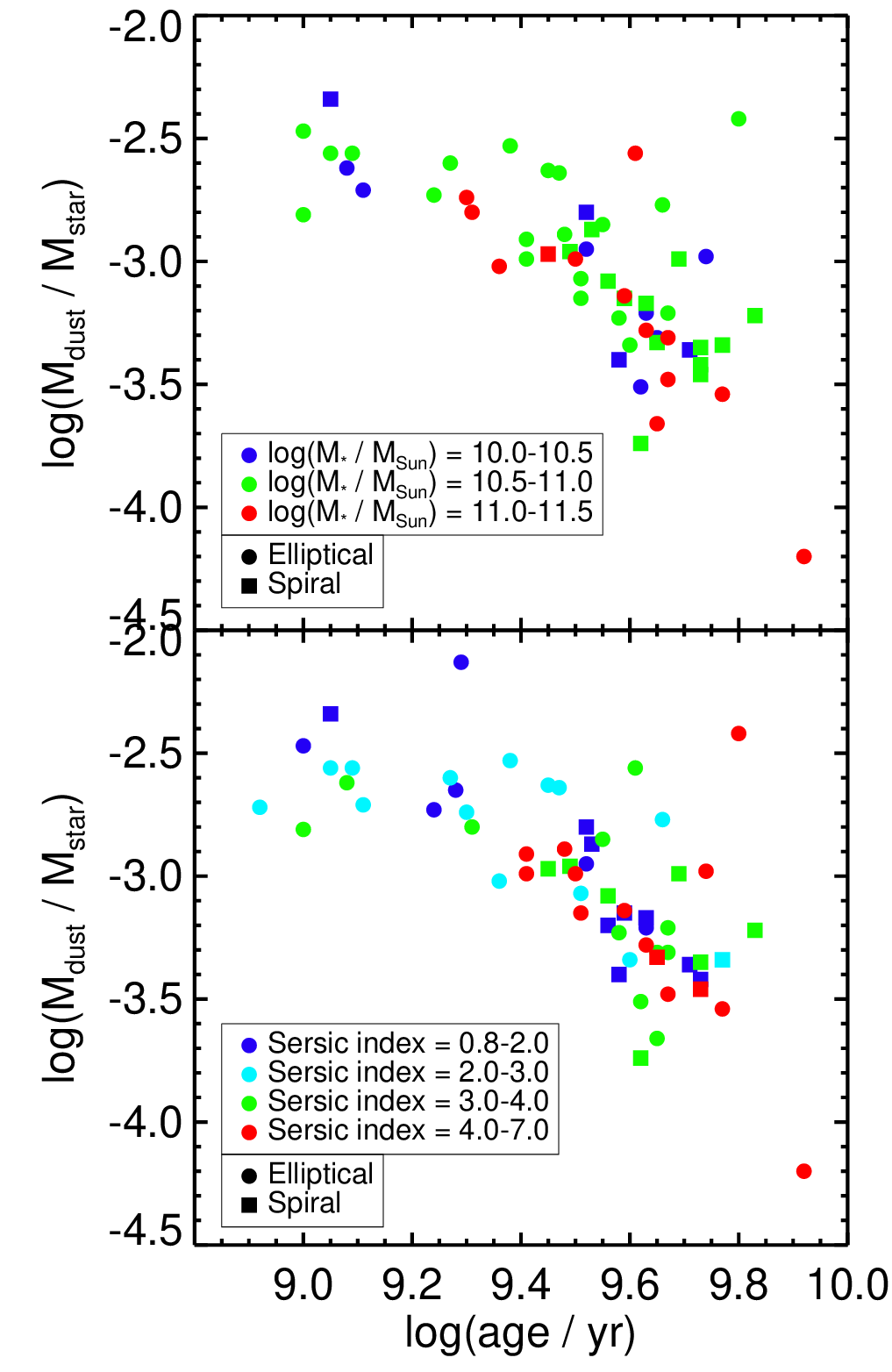}
 \caption{The same as Fig.~\ref{fig:mdms_age}, but with datapoints colour-coded by stellar mass ({\it top}) or Sersic index ({\it bottom}), as indicated in legend, and with symbols dependent on morphology (circles: ellipticals, {\it }squares:\ red spirals). The trend is independent of the mass, the Sersic index ranges, and morphology, so this shows that it is not driven by the range of these properties  observable across the galaxies in our sample. 
}
 \label{fig:mdms_age_msbin}
\end{figure}

In view of the lack of any indication of the existence of very cold dust in our sample (as judged by our SCUBA-2 sub-sample), we proceed under the assumption that the {\it Herschel}-derived dust masses are reliable.

We note that the dust mass on average appears to be lower at higher ages in Fig.~\ref{fig:md_age}. At the same time,
SFR and specific SFR are also lower on average, whereas stellar mass is somewhat higher. We do not detect any significant correlation between the age and the dust temperature or environment (Fig.~\ref{fig:all_age}), dust attenuation, optical size \citep{rowlands12}, or Sersic index (\citealt{kelvin12}; Fig.~\ref{fig:all_age2}), indicating that these properties are not driving dust evolution.
We note that dust attenuation in molecular clouds ($\tau_{\rm MC}$ in {\magphys}) is poorly constrained,
but in these galaxies molecular clouds contribute on average only $\sim25$\% of the infrared luminosity \citep{rowlands12}, so this parameter is not critical.

As indicated in
Fig.~\ref{fig:ms_age}, the stellar masses of the galaxies in our sample span three orders of magnitude, so
we use the dust mass relative to the stellar mass as an indicator of the dust content of a galaxy.
Hence, our analysis will be based on age trends of quantities normalised to the stellar mass. In this way, the evolutionary trends we describe are not affected by the fact that the galaxies in our sample are not equally massive. For example,
for galaxies with similar dust masses, but different stellar masses, we conclude that the more massive galaxy is more dust-deficient. This is true only in a mass-normalised sense.

In
Fig.~\ref{fig:mdms_age}, we plot the dust masses (normalised to the stellar mass of the galaxy) as a function
of age. It is evident that the dust-to-stellar mass ratio (a measure of the `dustiness' of a galaxy) decreases with stellar age. The typical uncertainties of these properties ($\sim0.3$ dex for the dust-to-stellar mass ratio $\mdust/\mstar$ and $\sim0.2$ dex for ages; \citealt{rowlands12})
are much smaller than the range of the values. 
The Spearman rank correlation coefficient is $-0.7$ with a very small probability ($\sim4\times10^{-11}$, $\sim5.5\sigma$, or $\sim5\times10^{-7}$, $\sim5\sigma$ without galaxies on the main-sequence) of the null hypothesis (no correlation) being acceptable.

We now assume that this can be interpreted as an evolutionary sequence (we discuss alternative explanations below). There is significant scatter in the relation, so we do not propose that individual galaxies can evolve into each other; rather, it should be understood in the sense that the progenitors of galaxies with higher ages drawn from our sample were broadly similar to those with lower ages: older galaxies have less dust because they have had more time to
get rid of it.

We can determine the timescale of the apparent dust decline by fitting
an exponential function 
\begin{equation}
\frac{\mdust}{\mstar}=Ae^{-{\rm age}/\tau},
\label{eq:exp}
\end{equation}
where $A$ is the normalisation constant and $\tau$ is the lifetime. We obtain a lifetime of $\tau=(2.5\pm0.4)\,$Gyr and a half-life time $t_{1/2}=\tau\ln(2)=(1.75\pm0.25)\,$Gyr. 
If we include only galaxies below the main sequence we get consistent  results: $\tau=(2.7\pm0.5)\,$Gyr, $t_{1/2}=(1.87\pm0.35)\,$Gyr.
This is the first time the dust removal timescale has been measured directly.

Interpreting the dust decline in Fig.~\ref{fig:mdms_age} 
as an evolutionary sequence, 
the fitting of equation (\ref{eq:exp}) gives an estimate of the dust removal timescale in early-type galaxies. This quantity may be used in models in which it needs to be assumed {\em a priori} and cannot be derived. Such a long timescale ($\sim2\,$Gyr) indicates that dust is present even in very old galaxies, as long as sufficient amounts have been formed in the past (see Sect.~\ref{sec:source} for a discussion on the source of dust in these galaxies).

Our estimate of the dust removal timescale is consistent with theoretical predictions of \citet{barlow78} and \citet{slavin15}. 
On the other hand, it is longer than the one measured by \citet[][$<45\pm25$\,Myr]{clemens10}. This is not surprising as their measurement is  based on the current AGB dust production rates, 
which leads to an underestimation of the lifetime if another dust production mechanism is dominant (e.g.
grain growth in the ISM).
Our derived lifetime of dust grains is much longer than that which is expected for a hot medium \citep{draine79,jones94,jones04}. Therefore, either there is not much hot gas in these galaxies, or dust is efficiently shielded in dense pockets. 
For a given stellar mass of an ETG there is spread in the  X-ray luminosity of the hot gas of more than 2 orders of magnitude  \citep{boroson11,su15} so it is not possible to predict whether galaxies in our sample contain such gas.

It is helpful to compare the dust removal timescale we derive with the quenching timescale of shutting down star formation.
\citet{peng15} and \citet{trussler19} measure a timescale for quenching by strangulation, a process involving halting gas inflows onto galaxies.
The timescale is 2--5\,Gyr for galaxies with masses $\log{M_*/\msun}=10$--$11$ independent of environment (cluster or field galaxies).
Similarly, simulated galaxies in the same mass range exhibit a quenching timescale of 2--4\,Gyr \citep{wright19}.
The dust removal lifetime we derive is similar or up to a factor of two shorter than these quenching timescales. 
This suggests that the shutdown of star formation and the decline in dust content may be connected.
It is, however, unlikely that the mechanism of the dust decline is directly related to star formation, for example, by astration or SN dust grain destruction. Fig.~\ref{fig:sfr_age} shows that the SFR is declining with age, so the effect of such processes is the weakest at high ages. Hence, if these processes were responsible for the dust decline, then the decline would be flattening at high ages, whereas Fig.~\ref{fig:mdms_age} demonstrates a steepening with age.
This effect is demonstrated in Fig.~\ref{fig:all_age}. We fit an exponential function to the SFR vs.~age panel and calculate the resulting removal of dust, assuming a conservatively low gas-to-dust mass ratio of 100. The evolution is too flat to explain the data and it is, indeed, flattening at high ages. A higher gas-to-dust mass ratio would result in an even flatter evolution.

In Fig.~\ref{fig:md_sfr}, we show how the dust decline we measure is related to the SFR decline (the manifestation of quenching). On the {\mdust}-SFR diagram, the dusty early-type galaxies are evolving away from the relation measured for star-forming galaxies \citep{dacunha10}, diagonally towards the bottom-left\footnote{See \citet{lianou16} for similar location of dusty ETGs but without the age information.}. Such behaviour has been predicted by \citet{hjorth14} to be a manifestation of a quenching process which does not require removing the ISM (otherwise a galaxy would evolve parallel to the {\mdust}-SFR relation). One possible quenching mechanism is morphological quenching, which means it would make the gas resilient against fragmentation by the influence of the bulge \citep{martig09,martig13,bitsakis19,lin19}. 
This evolution away from the {\mdust}-SFR relation is also evident in Fig.~\ref{fig:mdsfr_age}, which shows the distance from this relation as a function of age. The galaxies in our sample, as they leave the star-formation main-sequence, are also leaving the {\mdust}-SFR relation.

Finally, we note that the dust removal timescale we report here is measured using a sample of early-type and red spiral galaxies with detected dust emission, so there is no evidence that it applies to different galaxies. Star-forming galaxies are likely to have
quicker dust destruction processes, due to, for example, frequent SN explosions and more turbulent ISM.

The ratio of the dust mass to  stellar (or baryonic) mass has been used as an indicator of the evolutionary stage of  galaxies. The dust decline we detect is qualitatively consistent with  models of dust evolution by \citet{clark15}, \citet{calura17}, and  \citet{devis17,devis17b,devis19}. They conclude that the dust-to-baryon (or stellar) ratio first increases for very young galaxies and then declines in a similar way as is measured here. The evolution in some of these models is traced by the atomic gas fraction (from high to low) and indeed early-type galaxies with such measurements in these works have low gas mass fractions. 
On the other hand, the models of \citet{calura17} are difficult to compare with our study because they evolve the proto-spheroid version to less than 1\,Gyr, which is not considered here.

\citet{smercina18} interpret the dust decline in post-starburst systems as the result of sputtering due to the ISM being dominated by hot gas. The galaxies in our sample are older but it is possible that this effect is continuing to be important for them as well.

\subsection{Potential biases}

The dust decline in Fig.~\ref{fig:mdms_age} relies on the accuracy of the {\magphys} age measurement. Such measurements based on spectral energy distributions (SEDs)  are, indeed, shown to be reliable in assigning a galaxy to an age bin at least up to the age of 7\,Gyr  \citep{yi03}, which is the oldest age of the galaxies in our sample. Moreover, the age estimate we use correlates with the D4000 spectral break strength \citep{balogh99} from single fiber spectroscopy in the GAMA survey (Fig.~\ref{fig:d4000}; \citealt{gordon17}\footnote{\url{www.gama-survey.org/dr3/data/cat/SpecLineSFR/}}), which is an indicator of the age of the stellar population independent of the photometric {\magphys} estimate. This correlation has a Spearman rank correlation coefficient of $0.6$ with a very small probability of $\sim2.5\times10^{-7}$ ($\sim5.2\sigma$) of the null hypothesis (no correlation).
Finally, the age measurement we use correlates with the specific SFR ($\mbox{sSFR}\equiv\mbox{SFR}/\mstar$; Fig.~\ref{fig:ssfr_age}), which is a proxy for the inverse of the mass-weighted age, because the numerator (SFR) informs us about the rate at which young stars are being born, whereas the denominator (\mstar)  is usually dominated by older stellar populations. The significance of the sSFR-age correlation has a Spearman rank correlation coefficient of $-0.7$ with a very small probability of $\sim3.8\times10^{-10}$ ($\sim5.4\sigma$) of the null hypothesis (no correlation).

It is recognised that in SED modelling, the age and the stellar mass are to some degree degenerate (an older population has higher mass-to-light ratio, so it will result in higher stellar mass for the same fluxes). However, this effect does not drive the $\mdust/\mstar$ decline in Fig.~\ref{fig:mdms_age}, because this decline is mostly driven by the $
\mdust$ decline, not the $\mstar$ increase (see Fig.~\ref{fig:all_age}).

It is unlikely that the apparent evolution of ${\mdust}/M_*$ with age is caused by the selection effect. The sample selection was based on {\it Herschel} detections (dust-mass limit) and optical morphology. Therefore, if galaxies in the top-right corner of Fig.~\ref{fig:mdms_age} existed (old but dusty) they would be detected and included in the sample. 
Indeed, the dust mass detection limit for the H-ATLAS survey at the redshifts of our galaxies is  $\sim10^{5.2}\,\msun$ at $z=0.05$ and
$\sim10^{6.7}\,\msun$ at $z=0.3$ \footnote{Estimated from the GAMA {\magphys} catalogue including {\mdust} \citep{driver16,baldry18} keeping only sources with $250\,\micron$ fluxes greater than 32\,mJy, the limit used by \citet{rowlands12}; \url{www.gama-survey.org/dr3/data/cat/MagPhys}}, much lower than the lower envelope of the measurements for our sample (see Fig.~\ref{fig:all_age}).
Similarly, galaxies in the bottom-left corner, below the trend, would have detectable dust masses as they would be well above the detection limit.

Moreover, the trend in Fig.~\ref{fig:mdms_age} is not driven by the different stellar masses or morphologies of galaxies in our sample. In Fig.~\ref{fig:mdms_age_msbin}, we show this trend using different colours for three stellar mass bins, or for four Sersic index bins, along with different symbols for ellipticals and red spirals. It is evident that the trend persists even in narrower mass or Sersic index bins and restricting them to a single morphological type. In fact, the trend exhibits smaller scatter keeping only galaxies with Sersic indices larger than $4$, apart from two outliers.
Moreover, the Sersic index in our sample is not correlated with age (Fig.~\ref{fig:sersic}).

\begin{figure}
\includegraphics[width=0.48\textwidth]{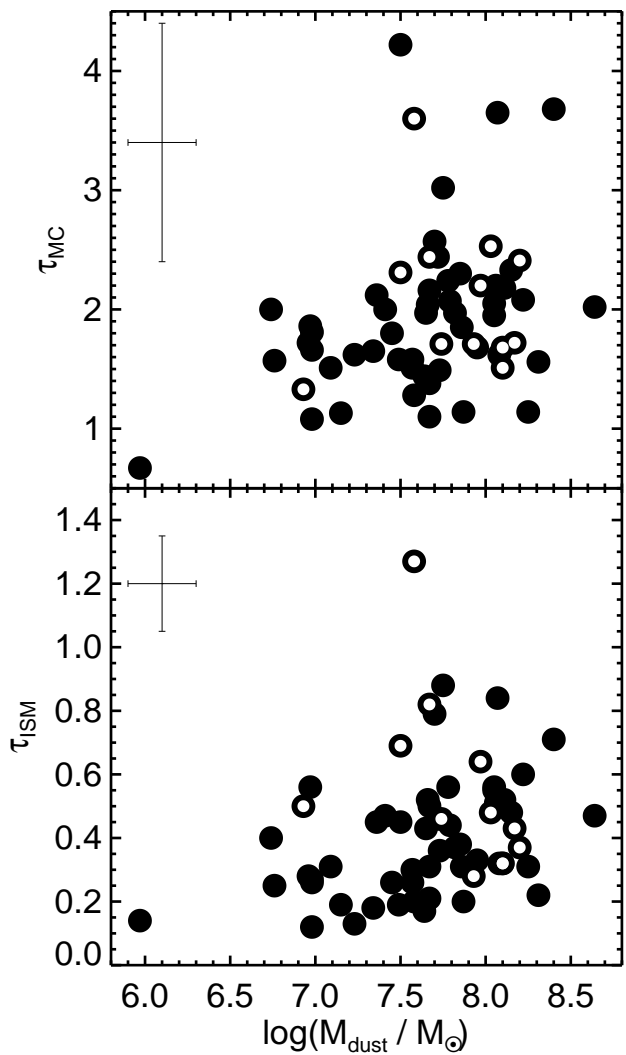}
 \caption{Optical depth in molecular clouds ({\it top}) and  diffuse ISM ({\it bottom}) as function of dust mass in the {\magphys} model derived by \citet{rowlands12}. 
As expected, there is a weak correlation between the stellar attenuation and dust mass.
}
\label{fig:all_md}
\end{figure}

The trend in Fig.~\ref{fig:mdms_age} is not a result of the age-attenuation degeneracy in the SED fitting because there is no trend in the derived ages and attenuation  (Fig.~\ref{fig:taumc}). If this were a problem then galaxies with high attenuation would be erroneously assigned high ages. However, these galaxies are the least dusty in our sample and are not more compact than the other members in the sample (see Fig.~\ref{fig:taumc}), so the stellar populations are unlikely to suffer the highest attenuation.
Additionally, consistent to the expectation, the stellar attenuation in the {\magphys} modelling is weakly correlated to the total dust masses in these galaxies (Fig.~\ref{fig:all_md}). The Spearman rank correlation coefficient is 0.36 with a probability of $\sim0.004$ ($\sim3\sigma$) of the null-hypothesis that there is no correlation.

Finally, our sample is unlikely to be affected by the age-metallicity degeneracy because most of these galaxies span the stellar gas range  of $\log(\mstar/\msun)=10.5$--$11.5$, at which the mass-metallicity relation is flat. Hence, the expected metallicities of our galaxies are within a narrow range of $12+(\mbox{O}/\mbox{H})=9.05$--$9.13$ \citep{tremonti04}.

\section{The source of dust in early-type galaxies}
\label{sec:source}

The origin of dust in galaxies in general, and in ellipticals in particular, is a matter of current debate.
While we can interpret the trend in Fig.~\ref{fig:mdms_age} as dust removal in action, we do not posses direct information about past dust evolution. 

In general, the source of dust can be internal or external. Internal dust formation mechanisms include AGB stars and SNe, with possible significant dust mass accumulation in the ISM. This means that the detected dust is what has remained from the pre-existing (before quenching) dust amount, or it is being formed by a current population of AGB stars. Alternatively, dust might have been recently brought in from an external source following a merging with dust-rich galaxies. 

We now demonstrate that the contribution of AGB stars to dust production is minor \citep[see also][]{rowlands12}
based on similar arguments as those presented in \citet{michalowski10qso,michalowski10smg4}, \citet{michalowski15}, and \citet{lesniewska19} for high-redshift galaxies
\citep[see also][]{morgan03,dwek07,dwek11,rowlands14b}.
Stellar masses of $10^{11}\,\msun$ imply around $10^{10}$ AGB stars ($1.5-8\,\msun$), assuming the \citet{chabrier03} IMF. 
In order to explain the measured dust masses of $\sim10^8\,\msun$ for galaxies in our sample at the lowest ages, each of the AGB stars would need to produce $\sim10^{-2}\,\msun$ of dust. This is the highest AGB dust yields to have been found theoretically \citep{morgan03,ferrarotti06,ventura12,nanni13,nanni14,schneider14}, which is valid only for narrow mass and metallicity ranges. On average, AGB stars produce less dust than this maximum by at least a factor of ten \citep{gall11c}. Similarly, \citet{rowlands12} also conclude that the detected dust masses in these galaxies cannot be explained by AGB dust production.
Hence, AGB stars could not have dominated the dust production in these galaxies.


These dust masses are larger by one order of magnitude than the prediction of recent SN dust production proposed by  \citet{gall18} for star-forming galaxies. This is expressed as $M_{\rm dust}^{\rm SN}=\mu_D\mbox{SFR}\Delta t$, where $M_{\rm dust}^{\rm SN}$ is the total dust mass formed by SNe in the current star-formation episode, $\mu_D$ (equal to 0.004 for the Chabrier IMF) is a product of the dust mass formed by one SN and the rate of SNe per unit SFR,  $\Delta t$ is the duration of the star-formation episode, for which we conservatively assume a large value of $1\,$Gyr. Indeed, it is unlikely that the dust detected for the galaxies in our sample was formed recently as the current SFRs are too low. The observed dust must have formed in the past, when SFRs were much higher. In a galaxy with a stellar mass of $10^{11}\,M_\odot$ with the \citet{chabrier03} IMF, there are $\sim1.2\times10^9$ stars with masses $8$--$40\,M_\odot$, which can explode as SNe. Therefore, one SN would need to produce $\sim0.8\,M_\odot$ of dust to explain the measured dust masses. This is close to the maximum SN dust yield with no dust destruction \citep{todini01,nozawa03,matsuura11,gall14,hjorth14,gall18}. Moreover, some stars with very high masses ended up collapsing into a black hole without producing dust dust all. Hence, SNe could be considered the source of dust if they were maximally efficient and did not destroy any dust.

ISM grain growth or very efficient SN dust production has also been invoked for this sample \citep{rowlands12} and for distant galaxies \citep[][]{dwek07,michalowski10smg4,michalowski10qso,michalowski15,gall11b, hirashita11,hjorth14,rowlands14b,valiante14,lesniewska19}, some of which may be representative progenitors of early-type galaxies in our sample. 
\citet{hirashita15} shows that grain growth in the ISM of ellipticals can operate on relatively short timescales of tens of Myr. That is fast enough to produce the considerable amounts of dust  detected in galaxies in our sample shortly after they were quenched.
On the other hand, \citet{ferrara16} shows that dust grain growth is too slow in the diffuse ISM, whereas in molecular clouds, this process has no influence on the total dust mass, because accreted material forms icy mantles which do not survive when a grain leaves the cloud.
Hence, the issue of grain growth is not yet understood.

Alternatively, dust in our galaxies could be of external origin, brought on by mergers with gas-rich galaxies, but the properties of the galaxies in our sample are inconsistent with this scenario. 
It predicts that $\mdust$ should be only weakly correlated with {\mstar} because the gas-rich companions would not bring in a significant amount of stars, so the final {\mstar} would  depend quite weakly on the number of merger events.
However, Fig.~\ref{fig:md_ms} shows that {\mdust} is correlated with {\mstar}.
The Spearman rank correlation coefficient is 0.5, with a very small probability ($\sim3\times10^{-5}$, $\sim4.2\sigma$, or $\sim5\times10^{-5}$, $\sim4.1\sigma$ without the galaxies on the main sequence) of the null hypothesis (no correlation) being acceptable.
This suggests that current dust reservoirs have been formed by the stellar populations present in these galaxies. In contrast \citet{diseregoalighieri13} finds no correlation between these quantities for dusty ellipticals in the Virgo cluster \citep[see also][]{davis11,smith12,kaviraj12,kaviraj13,dariush16,beeston18,kokusho19}, leading to the conclusion that the dust has been brought in by gas-rich mergers.

Moreover,
in the external origin scenario, {\mstar} should be independent of 
or slightly decreasing in proportion to
the derived age because  
galaxies with lower inferred age (and higher dust masses brought by mergers) have, on average, experienced more merging and, thus, they should be more massive. 
However, Fig.~\ref{fig:ms_age} shows that the stellar mass is increasing with age, which should not be the case if the apparent dust decline with age was a result of a merging with smaller galaxies at high derived ages and with larger galaxies at low ages.

Over several billion years, the size of a galaxy is also expected to grow as a result of minor mergers \citep{naab09,hopkins10b,trujillo11,furlong17}. This is not observed for our sample (Fig.~\ref{fig:size}).

Finally, the incidence of high dust masses of $\sim10^8\,\msun$ at lower stellar ages makes it unlikely that most of the dust has been brought in by minor mergers.
Indeed, \citet{rowlands12}  found unfeasibly high number of mergers compared to simulations of what is needed to supply enough dust. Additionally, they did not find morphological disturbances for these galaxies, which would be expected in cases of mergers.

We also note that \citet{griffith19} present evidence of internal production of dust for three other ellipticals by finding that their stellar and gas metallicities are similar, which would not be the case if dust was brought by a merger with a low-metallicity dwarf. Moreover \citet{babyk19} find that the molecular content of ellipticals is correlated with their hot gas reservoir, 
which also suggests an internal origin of the molecular gas. 
Finally, \citet{bassett17} find the kinematics of dusty early-type galaxies consistent with recent minor mergers for only 8\% of their sample. 
Similarly, in a small sample of early-type galaxies, \citet{sansom19} find carbon monoxide (CO) emission line rotational axes aligned with photometric axes, which is inconsistent with a recent merger/inflow hypothesis.

We conclude that it is likely that the dust in galaxies in our sample is of internal origin. It was either formed in the past directly by SNe (when SFRs were high) or with seeds produced by SNe, and with grain growth in the ISM significantly contributing to the dust mass accumulation shortly after the galaxies were quenched.

\section{Conclusions}

We have analysed the dust properties of a sample of galaxies detected by {\it Herschel} in the far-infrared based on the early-type morphology or spirals with red colours identified visually. We have found no obvious biases which could affect our analysis. We discovered an exponential decline of the dust-to-stellar mass ratio with age, which we interpret as an evolutionary trend of dust removal from these galaxies.
For the first time, we have directly measured the dust removal timescale in such galaxies to be $\tau=(2.5\pm0.4)\,$Gyr (the corresponding half-life time is $\sim(1.75\pm0.25)$ Gyr).
This quantity may be used in models in which it needs to be assumed {\em a priori} and cannot be derived.
Any process which removes dust in these galaxies cannot, therefore, take place  on shorter timescales.

This timescale is comparable to the quenching timescales found in simulations for galaxies with similar stellar masses.  
This suggests that the shutdown of star formation and the decline in dust content may be connected.

We find that the dust in these galaxies is likely of internal, not external origin. It was either formed in the past directly by SNe or with seeds produced by SNe, and with grain growth in the ISM significantly contributing to the dust mass accumulation.

In future papers we intend to constrain the physical mechanism of this process based on the gas properties of these galaxies and theoretical modelling.

\label{sec:conclusion}

\begin{acknowledgements}

We thank our anonymous referee and Joanna Baradziej
for help with improving this paper, 
Sune Toft, Georgios Magdis and Johan Fynbo for useful discussions,
 Harriet Parsons for the help with the JCMT observations,
and JCMT staff and observers for executing our programs.
 
M.J.M.~acknowledges the support of 
the National Science Centre, Poland through the POLONEZ grant 2015/19/P/ST9/04010 and SONATA BIS grant 2018/30/E/ST9/00208,
the UK Science and Technology Facilities Council,
British Council Researcher Links Travel Grant, Royal Society of Edinburgh  International Exchange Programme  and the hospitality of the Instituto Nacional de Astrof\'{i}sica, \'{O}ptica y Electr\'{o}nica (INAOE), the Academia Sinica Institute of Astronomy and Astrophysics (ASIAA), and the DARK.
This project has received funding from the European Union's Horizon 2020 research and innovation programme under the Marie Sk{\l}odowska-Curie grant agreement No. 665778.
J.H. was supported by a VILLUM FONDEN Investigator grant (project number 16599).
C.G acknowledges funding by the Carlsberg Foundation and the Carnegie Trust Research Incentive Grant (PI: M.~Micha{\l}owski).
H.H.~thanks the Ministry of Science and Technology for support through grant MOST 105-2112-M-001-027-MY3 and MOST 107-2923-M-001-003-MY3 (RFBR 18-52-52-006).
This work has been supported by the Grant-in-Aid for
Scientific Research (No. 17H01110). 

The James Clerk Maxwell Telescope is operated by the East Asian Observatory on behalf of The National Astronomical Observatory of Japan, Academia Sinica Institute of Astronomy and Astrophysics, the Korea Astronomy and Space Science Institute, the National Astronomical Observatories of China and the Chinese Academy of Sciences (Grant No. XDB09000000), with additional funding support from the Science and Technology Facilities Council of the United Kingdom and participating universities in the United Kingdom and Canada. 
Additional funds for the construction of SCUBA-2 were provided by the Canada Foundation for Innovation. 
This research used the facilities of the Canadian Astronomy Data Centre operated by the National Research Council of Canada with the support of the Canadian Space Agency. 
GAMA is a joint European-Australasian project based around a spectroscopic campaign using the Anglo-Australian Telescope. The GAMA input catalogue is based on data taken from the Sloan Digital Sky Survey and the UKIRT Infrared Deep Sky Survey. Complementary imaging of the GAMA regions is being obtained by a number of independent survey programmes including GALEX MIS, VST KiDS, VISTA VIKING, WISE, Herschel-ATLAS, GMRT and ASKAP providing UV to radio coverage. GAMA is funded by the STFC (UK), the ARC (Australia), the AAO, and the participating institutions. The GAMA website is http://www.gama-survey.org/ .

This research has made use of 
the Tool for OPerations on Catalogues And Tables \citep[TOPCAT;][]{topcat}: \url{www.starlink.ac.uk/topcat/ };
SAOImage DS9, developed by Smithsonian Astrophysical Observatory \citep{ds9};
the NASA's Astrophysics Data System Bibliographic Services
and the WebPlotDigitizer\footnote{\url{https://automeris.io/WebPlotDigitizer}} of Ankit Rohatgi.

\end{acknowledgements}




\end{document}